\documentclass[PDFLaTeX, twocolumn]{article}

\usepackage[affil-it]{authblk}	
\usepackage{amsmath} 		
\usepackage{amssymb}  		
\usepackage{mathtools}
\usepackage{subfig} 			
\usepackage{multirow}  		
\usepackage{relsize}
\usepackage{booktabs}		
\usepackage{url}		
\usepackage{caption}

\renewcommand{\vec}[1]{\mathbf{#1}}

\newcommand{\norm}[1]{\left\lVert #1 \right\rVert}

\newtheorem{definition}{Definition}%

\author[1]{Andrea Giusti \footnote{These authors contributed equally to this work.}}
\author[2]{Gian Carlo Maffettone  {\small*}}
\author[3]{Davide Fiore}
\author[2]{Marco Coraggio}
\author[1,2]{Mario di Bernardo \thanks{Corresponding Author}}

\affil[1]{{\small Department of Electrical Engineering and Information Technology, University of Naples Federico II, Via Claudio 21, Naples, 80125, Italy}}
\affil[2]{{\small Scuola Superiore Meridionale, School for Advanced Studies, Largo S. Marcellino, 10, Naples, 80138, Italy}}
\affil[3]{{\small Department of Mathematics and Applications “R. Caccioppoli”, University of Naples Federico II, Via Cintia, Monte S.Angelo, Naples, 80126, Italy}}

\title{Distributed control for geometric pattern formation of large-scale multirobot systems}

\date{}

\begin{document}

\maketitle



\abstract{Geometric pattern formation is crucial in many tasks involving large-scale multi-agent systems. Examples include mobile agents performing surveillance, swarm of drones or robots, or smart transportation systems.
Currently, most control strategies proposed to achieve pattern formation in network systems either show good performance but require expensive sensors and communication devices, or have lesser sensor requirements but behave more poorly. Also, they often require certain prescribed structural interconnections between the agents (e.g., regular lattices, all-to-all networks etc).
In this paper, we provide a distributed displacement-based control law that allows large group of agents to achieve triangular and square lattices, with low sensor requirements and without needing communication between the agents. Also, a simple, yet powerful, adaptation law is proposed to automatically tune the control gains in order to reduce the design effort, while improving robustness and flexibility.
We show the validity and robustness of our approach via numerical simulations and experiments, comparing it with other approaches from the existing literature.}

\maketitle

\section{Introduction}
\label{sec::intro}

\subsection{Problem description and motivation}

Many robotic applications require---or may benefit from---one or more groups of multiple agents to perform a joint task \cite{Dudek1996}; this is, for example, the case of surveillance, exploration, herding \cite{Auletta2022} or transportation \cite{Bayindir2016a}.
When the number of agents becomes extremely large, the task becomes a \emph{swarm robotics} problem \cite{Brambilla2013}. 
Typically, in these problems, it is assumed that the agents are relatively simple, and thus have limited communication and sensing capabilities, and limited computational resources; see for example the robotic swarms described in \cite{Hauert2009a, Rubenstein2014a, Gardi2022}.

In swarm robotics, typical tasks of interest include \emph{aggregation}, \emph{flocking}, \emph{foraging}, \emph{object clustering}, \emph{navigation}, \emph{spatial organisation}, \emph{collaborative manipulation}, and \emph{task allocation} \cite{Brambilla2013,Bayindir2016a}. 
Among these, an important subclass of spatial organisation problems is \emph{geometric pattern formation}, where the goal is for the agents to self-organize their relative positions into some desired structure or {\em pattern}, e.g., multiple adjacent triangles.
Pattern formation is crucial in many applications \cite{Oh2017a}, including sensor networks deployment \cite{Zhao2019, Kim2014}, collective search and rescue \cite{Wong2005, Pugh2007}, collective transportation and construction \cite{Rubenstein2013, Mooney2014}, and 3D-2D exploration and mapping \cite{Thrun2000}. 
%
There are two main difficulties associated with achieving pattern formation. 
Firstly, as there are no leader agents, the pattern must emerge by exploiting a control strategy that is the same for all agents, \emph{distributed} and \emph{local} (i.e., each agent can only use information about ``nearby'' agents).
Secondly, the number of agents is large and may change over time; therefore the control strategy must also be \emph{robust} to uncertainties in the size of the swarm and to its possible variations.

This sets the problem of achieving pattern formation apart from the more classical \emph{formation control} problems \cite{Oh2015} where agents are typically fewer and have pre-assigned roles within the formation.

Nevertheless, some of the theory and solutions developed for formation control may be exploited to describe pattern formation. For this reason, to classify existing solutions to pattern formation, we employ the same taxonomy proposed in \cite{Oh2015} for formation control, which is based on the type of information available to the agents.
Namely, existing strategies can be classified as being (i) \emph{position-based} when it is assumed agents know their position and orientation and those of their neighbours, in a global reference frame;
(ii) \emph{displacement-based} when agents can only sense their own orientation with respect to a global reference direction (e.g., North) and the relative positions of their neighbours;
(iii) \emph{distance-based} when agents can measure the relative positions of their neighbours with respect to their local reference frame. 
In terms of sensor requirements, position-based solutions are the most demanding, requiring global positioning sensors, typically GPS, and communication devices, such as WiFi or LoRa. 
Differently, displacement-based methods require only a distance sensor (e.g., LiDAR) and a compass, although the latter can be replaced by a coordinated initialisation procedure of all local reference frames \cite{Cortes2009}.
Finally, distance-based algorithms are the less demanding, needing only some distance sensors.

A pressing open challenge in pattern formation problems is devising new control strategies that can combine low sensor requirements with high and consistent performance.
This is crucial in swarm robotics, where it would be cumbersome or prohibitively expensive to equip all agents with GPS sensors and communication capabilities.

\subsection{Related work}

\subsubsection*{Position-based approaches}

In \cite{Pinciroli2008}, a position-based algorithm was proposed to achieve 2D triangular lattices in a constellation of satellites in a 3D space.
This strategy combines global attraction towards a reference point with local interaction among the agents to control both the global shape and the internal lattice structure of the swarm.
In \cite{Casteigts2012}, a position-based approach was presented that combines the common radial virtual force (also used in \cite{Spears1999,Spears2004,Hettiarachchi2005,Torquato2009}) with a normal force. 
In this way, a network of connections is built such that each agent has at least two neighbours; then, a set of geometric rules is used to decide whether any or both of these forces are applied between any pair of agents.
Importantly, this approach requires the acquisition of positions from two-hop neighbours.
In \cite{Zhao2019}, a position-based strategy is presented to achieve triangular and square patterns, as well as lines and circles, both in 2D and 3D; the control strategy features global attraction towards a reference point and re-scaling of distances between neighbours, with the virtual forces changing according to the goal pattern.
A qualitative comparison was also provided with the distance-based strategy from \cite{Spears2004}, showing more precise configurations and a shorter convergence time, due to the position-based nature of the solution. 

\subsubsection*{Displacement-based approaches}

In \cite{Li2009}, a displacement-based approach is presented based on the use of a geometric control law similar to the one proposed in \cite{Lee2008}. The aim is to obtain triangular lattices but small persisting oscillations of the agents are present at steady state, as the robots are assumed to have a constant non-zero speed.
In \cite{Balch2000, Balch2000a}, an approach is discussed inspired by covalent bonds in crystals, where each agent has multiple attachment points for its neighbours. 
Only starting conditions close to the desired pattern are tested, as the focus is on navigation in environments with obstacles.
Finally, in \cite{Song2014} the desired lattice is encoded by a graph, where the vertices denote possible \emph{roles} the agents may play in the lattice and edges denote rigid body transformations between the local frames or reference of pairs of neighbours.
All agents communicate with each other and are assigned a label (or identification number) through which they are organised hierarchically to form triangular, square, hexagonal or octagon-square patterns.

\subsubsection*{Distance-based approaches}
A popular distance-based approach for the formation of triangular and square lattices, named \emph{physicomimetics}, was proposed in \cite{Spears1999} and later also studied in \cite{Spears2004, Hettiarachchi2005}. 
The control strategy is based on the use of \emph{virtual forces} \cite{Khatib1985}, an approach inspired by Physics, where each agent is subject to virtual forces (e.g., Lennard-Jones and Morse functions \cite{Brambilla2013, DOrsogna2006}) from neighbouring agents, obstacles, and the environment.
In these studies (\cite{Spears1999, Spears2004, Hettiarachchi2005}), triangular lattices are achieved with long-range attraction and short-range repulsion forces only, while square lattices are obtained through a selective rescaling of the distances between some of the agents.
An extension for the formation of hexagonal lattices was proposed in \cite{Sailesh2014}, but with the requirement of an ad hoc correction procedure to prevent agents from remaining stuck in the centre of a hexagon. 
The main drawback of the physicomimetics strategy (\cite{Spears1999, Spears2004, Hettiarachchi2005, Sailesh2014}) is that it can produce the formation of multiple aggregations of agents, each respecting the desired pattern, but with different orientations.
Another problem, described in \cite{Spears2004}, is that, for some values of the parameters, multiple agents can converge towards the same position and collide.
In \cite{Torquato2009}, an approach exploiting Lennard-Jones-like virtual forces is numerically optimised to stabilise locally a hexagonal lattice.
When applied to mobile agents, the interaction law is time-varying and requires synchronous clocks among the agents.
In \cite{Lee2008}, a different distance-based control strategy, derived from geometric arguments, was proposed to achieve the formation of triangular lattices. 
In this study, an analytical proof of convergence to the desired lattice is given exploiting Lyapunov methods.
Robustness to agents' failure and the capability of detecting and repairing holes and gaps in the lattice are obtained via an ad hoc procedure and verified numerically.
A 3D extension was later presented in \cite{Lee2010}.

\subsection{Contribution}

In this paper, we introduce a \emph{distributed} \emph{displacement-based} control strategy to solve pattern formation problems in swarm robotics that requires no communication among the agents or labelling them.
In particular, to achieve triangular and square lattices we employ two virtual forces controlling the norm and the angle of their relative position, respectively.
The main contributions can be listed as follows 
\begin{enumerate}
    
    \item Our strategy performs significantly better than other distance-based algorithms (\cite{Spears1999,Sailesh2014}) when achieving square lattices, in terms of precision and robustness, with only a minimal increase in sensor requirements (a compass) and without needing the more costly sensors and communication devices used for position-based strategies.
    \item The control gains can be set automatically, according to a simple adaptive law, in order for the agents to organize themselves and switch from one pattern to the other.
    \item Numerical simulations and experiments show its effectiveness even in the presence of actuator constraints and other more realistic effects.
\end{enumerate}

\section{Preliminaries}
\label{sec::MathPreliminaries}
\subsubsection*{Notation}
We denote by $\Vert \cdot \Vert$ the Euclidean norm. 
Given a set $\mathcal{B}$, its cardinality is denoted by $\vert \mathcal{B} \vert$.
We refer to $\mathbb{R}^2$ as the \emph{plane}.

\subsection{Planar swarms}
\begin{definition}[Swarm]
\label{def:swarm}
A \emph{(planar) swarm} $\mathcal{S} \coloneqq \{1,2,\dots,N\}$ is a set of $N \in \mathbb{N}_{>0}$ identical agents that can move on the plane.
For each agent $i \in \mathcal{S}$, $\vec{x}_i(t)\in \mathbb{R}^2$ denotes its position in the plane at time $t \in \mathbb{R}$.
\label{def::swarm}
\end{definition}

\begin{figure}[t!]
    \centering
    \includegraphics[width=0.60\columnwidth]{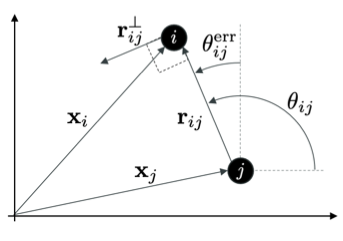}
    \caption{Geometrical relationship between a pair of agents.}
    \label{fig::vectors}
\end{figure}

Moreover, $\vec{r}_{ij}(t) \coloneqq \vec{x}_{i}(t)-\vec{x}_{j}(t) \in \mathbb{R}^2$ is the relative position of agent $i$ with respect to agent $j$, and $\theta_{ij}(t) \in [0, 2\pi]$ is the angle between $\vec{r}_{ij}$ and the horizontal axis (see Fig.~\ref{fig::vectors}). 

\begin{definition}[Neighbourhood]
\label{def::neigh}
Given a swarm and a \emph{sensing radius} $R_{\text{s}} \in \mathbb{R}_{>0}$, the \emph{neighbourhood} of agent $i$ at time $t$ is 
\begin{align}\label{eq:neighbourhood}
    \mathcal{N}_i(t) & \coloneqq \{ j \in \mathcal{S} \setminus \{i\} : \Vert \vec{r}_{ij}(t)\Vert \leq R_{\text{s}} \}.
\end{align}
\end{definition}

\begin{definition}[Adjacency set] 
\label{def::adjacency_set}
Given a swarm and some finite $R_{\min}, R_{\max} \in \mathbb{R}_{>0}$, with $R_{\min}\leq R_{\max}$, the \emph{adjacency set} of agent $i$ at time $t$ is (see Fig.~\ref{fig::lattices})
\begin{align}
    \mathcal{A}_i(t) &\coloneqq \{j \in \mathcal{S} \setminus \{i\} : R_{\min} \leq \Vert \vec{r}_{ij}(t)\Vert \leq R_{\max} \}.
\end{align}
\end{definition}

Notice that if $R_{\max} \leq R_{\text{s}}$ then $\mathcal{A}_i \subseteq \mathcal{N}_i$.

\begin{definition}[Links]
A \emph{link} is a pair $(i,j) \in \mathcal{S} \times \mathcal{S}$ such that $j \in \mathcal{A}_i(t)$ (or equivalently $i \in \mathcal{A}_j(t)$).
Moreover, $\mathcal{E}(t)$ is the set of all links existing at time $t$.
\end{definition}

Clearly, it is possible to associate to the swarm a time-varying graph $\mathcal{G}(t) = (\mathcal{S}, \mathcal{E}(t))$ \cite{Latora}; $\mathcal{S}$ and $\mathcal{E}(t)$ being the set of vertices and edges, respectively.\footnote{Formally, $\mathcal{G}(t)$ is a directed graph, even though $\mathcal{E}(t)$ is such that the existence of $(i, j)$ implies the existence of $(j, i)$.}

Finally, given any two links $(i,j)$ and $(h,k)$, we denote with $\theta_{ij}^{hk}(t) \in [0, 2\pi]$ the absolute value of the angle between the vectors $\vec{r}_{ij}$ and $\vec{r}_{hk}$.

\subsection{Lattice and performance metrics}

\begin{figure}[t!]
    \centering
    \includegraphics[width=1\columnwidth]{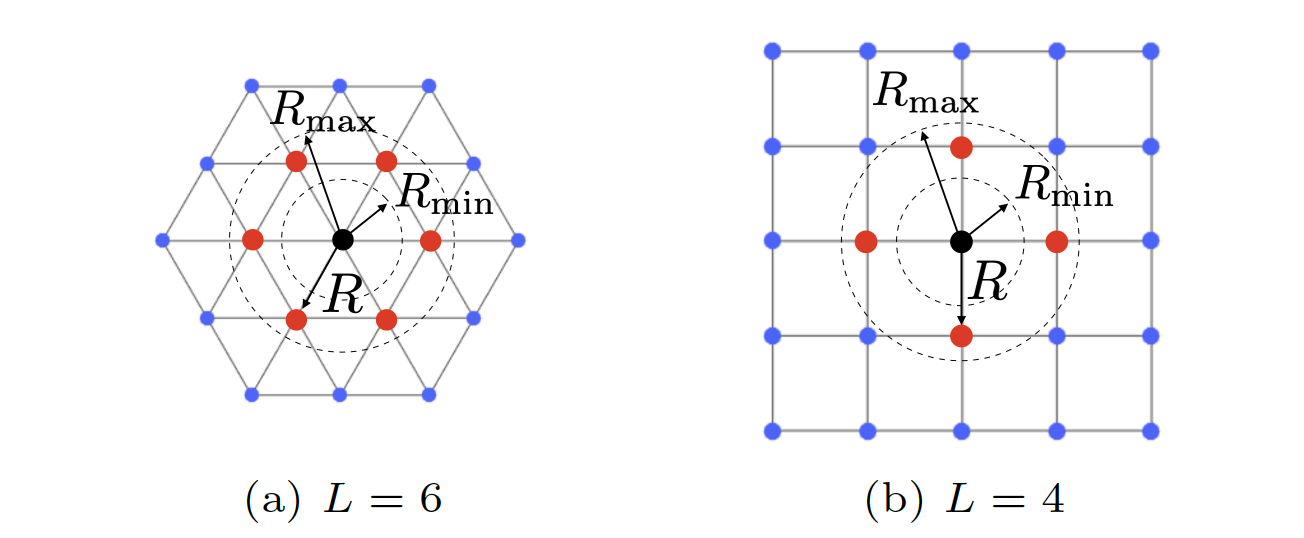}
    \caption{$(L,R)$-lattice formations for (a) triangular ($L=6$) and (b) square ($L=4$) lattices.
    Red dots are agents in the adjacency set ($\mathcal{A}_i$) of the black agent ($i$).
    }
    \label{fig::lattices}
\end{figure}

\begin{definition}[Lattice]
\label{def:lattice}
Given some $L \in \{4, 6\}$ and $R \in \mathbb{R}_{>0}$, a \emph{$(L, R)$-lattice} is a set of points in the plane that coincide with the vertices of an associated \emph{regular tiling} \cite{Grunbaum1977,Engel2004}; $R$ is the distance between adjacent vertices and $L$ is the number of adjacent vertices each point has.
\end{definition}
In Definition \ref{def:lattice}, $L=4$, and $L=6$ correspond to square and triangular lattices, respectively, as portrayed in Fig.~\ref{fig::lattices}.
We say that a swarm \emph{self-organises into a $(L,R)$-lattice} if (i) each agent has at most $L$ links, and (ii) $\forall (i,j) \in \mathcal{E}$ and $\forall (h,k) \in \mathcal{E}$ it holds that $\theta_{ij}^{hk}$ is some multiple of $2\pi/L$.
To assess whether a swarm self-organises into some desired $(L,R)$-lattice, we introduce two metrics.

\begin{definition}[Regularity metric]
\label{def::regularity}
Given a swarm and a desired $(L,R)$-lattice, the \emph{regularity metric} $e_{\theta}(t) \in [0,1]$ is
\begin{align}
    e_{\theta}(t) \coloneqq \frac{L}{\pi}\cdot \theta_{\mathrm{err}}(t),
    \label{eq:regMet}
\end{align}
where, omitting the dependence on time,
\begin{align}
    \theta_{\mathrm{err}} &\coloneqq\frac{1}{\vert \mathcal{E}\vert^2-2\vert \mathcal{E}\vert} \mathlarger{\sum}_{(i,j)\in \mathcal{E}} \mathlarger{\sum}_{(h,k) \in \mathcal{E}} \min_{q\in \mathbb{Z}} \left\vert\theta_{ij}^{hk} -q\frac{2\pi}{L}\right\vert .
    \label{eq::SpearsMetrics}
\end{align}
\end{definition}

The regularity metric $e_{\theta}$, derived from \cite{Spears1999}, quantifies the incoherence in the orientation of the links in the swarm.
In particular, $e_{\theta}=0$ when all the pairs of links form angles that are multiples of $2\pi/L$ (which is desirable to achieve the $(L, R)$-lattice), while $e_{\theta}=1$ when all pairs of links have the maximum possible orientation error, equal to $\pi/L$. 
Finally, $e_{\theta} \approx 0.5$ generally corresponds to the agents being arranged randomly.

\begin{definition}[Compactness metric]
\label{def::compactness}
Given a swarm and a desired $(L,R)$-lattice, the \emph{compactness metric} $e_L(t) \in [0,(N-1-L)/L]$ 
is
\begin{align}
    e_L(t)&\coloneqq\frac{1}{N} \sum_{i=1}^N \frac{\big\lvert \vert \mathcal{A}_i (t)\vert - L \big\rvert}{L}.
    \label{eq::compactness_metric}
\end{align}
\end{definition}
The compactness metric $e_L$ measures the average difference between the number of neighbours each agent has and the one they are ought to have in a $(L,R)$-lattice.
$e_L$ is maximum ($e_L = (N-1-L)/L$) when all agents are concentrated in a small region, and links exist between all pairs of agents.
$e_L=1$ when all the agents are scattered loosely in the plane, and no links exist between them.
Finally, $e_L = 0$ when all the agents have $L$ links (which is desirable to achieve the $(L, R)$-lattice).
It is important to remark that, if the number $N$ of agents is finite, $e_L$ can never be equal to zero, because the agents on the boundary of the group will always have less than $L$ links (Fig.~\ref{fig::lattices}).
This effect gets less relevant as $N$ increases.
Note that a similar metric is also independently defined in \cite{Song2014}.

For the sake of brevity, in what follows we will omit dependence on time when that is clear from the context.

\section{Control design}

\subsection{Problem formulation}
\label{sec::ProblemStatement}
Consider a planar swarm $\mathcal{S}$  whose agents' dynamics is described by the first order model
\begin{align}
    \dot{\vec{x}}_i(t) = \vec{u}_i(t), \ \ \forall \ i\in\mathcal{S},
    \label{eq::firstOrdDynamics}
\end{align}
where $\vec{x}_i(t)$ was given in Definition \ref{def:swarm} and $\vec{u}_i(t) \in \mathbb{R}^2$ is some input signal determining the velocity of agent $i$.%
\footnote{
First order models like \eqref{eq::firstOrdDynamics} are often used in the literature \cite{Lee2008, Lee2010, Casteigts2012, Zhao2019}.
In some other works \cite{Spears1999, Spears2004, Sailesh2014} a second order model is used, given by $m \ddot{\vec{x}}_i + \mu \dot{\vec{x}}_i = \vec{u}_i$, where $\vec{u}_i$ is a force, $m$ is a mass and $\mu$ is a viscous friction coefficient.
Under the simplifying assumptions of small inertia ($m\Vert \dot{\vec{v}}_i \Vert \ll \mu \Vert \vec{v}_i \Vert $) and $\mu = 1$, the two models coincide.}
We aim to solve the following control problem.
\paragraph*{Problem statement}
Design some \emph{distributed} feedback control law  $\vec{u}_i = \vec{g}(\{\vec{r}_{ij}\}_{j\in \mathcal{N}_i}, L, R)$
to let the swarm self-organise into a desired triangular or square lattice, starting from any set of initial positions in some disk of radius $r$. 
Moreover, we require the law to be:
\begin{enumerate}
    \item \emph{robust} to failures of agents and to noise;
    \item \emph{flexible}, allowing dynamic reorganisation into different patterns; 
    \item \emph{scalable}, allowing the number of agents $N$ to change dynamically.
\end{enumerate}
To assess the self-organising capability of the swarm, we seek to minimise the performance metrics $e_{\theta}$ and $e_L$ (see Definitions \ref{def::regularity} and \ref{def::compactness}).

\subsection{Distributed control law}
\label{sec::controlDesign}
Next, we present a distributed displacement-based control law that solves the problem described in Sec.~\ref{sec::ProblemStatement}.
Namely, we set
\begin{align}\label{eq::controlLaw}
    \vec{u}_i(t) &= \vec{u}_{\text{r},i}(t)+ \vec{u}_{\text{n},i}(t),
\end{align}
where $\vec{u}_{\text{r},i}$ and $\vec{u}_{\text{n},i}$ are the \emph{radial} and \emph{normal} control inputs, respectively.
The two inputs have different purposes and each comprises  several \emph{virtual forces}.
The radial input $\vec{u}_{\text{r},i}$ is the sum of attracting/repelling actions between the agents, with the purpose of aggregating them into a compact swarm, while avoiding collisions.
The normal input $\vec{u}_{\text{n},i}$ is also the sum of multiple actions, used to adjust the angles of the relative positions of the agents.

Law \eqref{eq::controlLaw} is \emph{displacement-based} because it only requires that each agent $i$ (i) can measure the relative positions of the agents close to it (in the sets $\mathcal{N}_i$ and $\mathcal{A}_i$), and (ii) has the knowledge of a common reference direction.
Next, we describe in detail the two control actions in \eqref{eq::controlLaw}.

\subsection{Radial Interaction}
\label{subsec::RIF}
The radial control input $\vec{u}_{\text{r},i}$ in \eqref{eq::controlLaw} is defined as the sum of several virtual forces, one for each agent in $\mathcal{N}_i$ (neighbours of $i$), each force being attractive (if the neighbour is far) or repulsive (if the neighbour is close).
Specifically,
\begin{align}\label{eq::radInput}
    \vec{u}_{\text{r},i} &= G_{\text{r},i}\sum_{j\in \mathcal{N}_i} f_{\text{r}}(\Vert \vec{r}_{ij}\Vert) \frac{\vec{r}_{ij}}{\Vert \vec{r}_{ij} \Vert},
\end{align}
where $G_{\text{r},i}\in \mathbb{R}_{\geq0}$ is the radial control gain.
Note that $\vec{u}_{\text{r},i}$ is termed as \emph{radial} input because in \eqref{eq::radInput} the attraction/repulsion forces are parallel to the vectors $\mathbf{r}_{ij}$ (see Fig. \ref{fig::vectors}).
The magnitude and sign of each of these actions depend on the distance ($\Vert \vec{r}_{ij}\Vert$) between the corresponding agents, according to the \emph{radial interaction function} $f_{\text{r}} : \mathbb{R}_{\geq 0} \rightarrow \mathbb{R}$.
Here, we select $f_{\text{r}}$  as the Physics-inspired Lennard-Jones function \cite{Brambilla2013, Hettiarachchi2005}, given by
\begin{equation}
    f_{\text{r}}(\Vert \vec{r}_{ij} \Vert) = \min \left\lbrace \left( \frac{a}{\Vert \vec{r}_{ij} \Vert^{2c}}-\frac{b}{\Vert \vec{r}_{ij} \Vert^c}\right), \ 1 \right\rbrace,
    \label{eq::Lennard-Jones}
\end{equation}
where $a, b \in \mathbb{R}_{> 0}$ and $c \in \mathbb{N}$ are design parameters.
In \eqref{eq::Lennard-Jones}, $f_{\text{r}}$ is saturated to $1$ to avoid divergence for $\lVert \vec{r}_{ij} \rVert \to 0$.
$f_\mathrm{r}$ is portrayed in Fig. 3a.

\begin{figure}[t!]
    \centering
    \includegraphics[width=1\columnwidth]{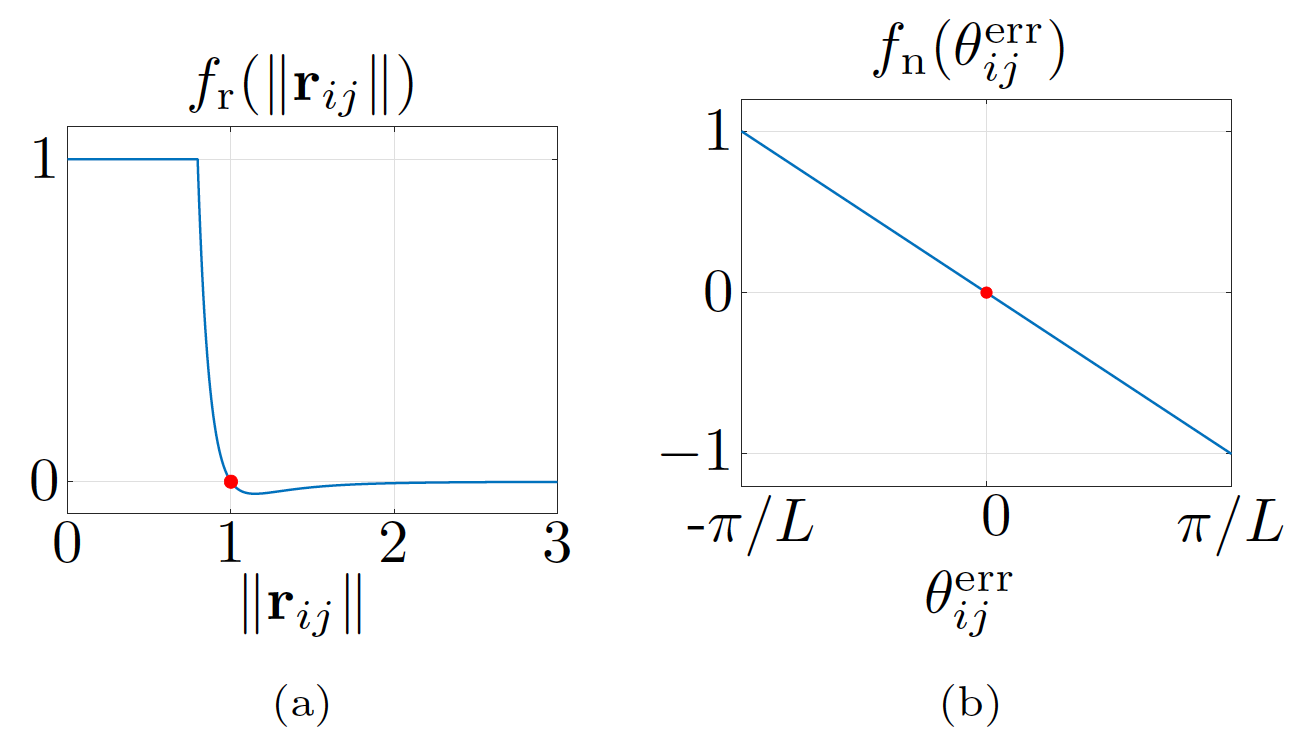}
    \caption{(a) Radial and (b) normal interaction functions. 
    Red dots highlight zeros of the functions.
    Parameters are taken from Tab \ref{tab:parameters}.    }
    \label{fig::forces}
\end{figure}

\subsection{Normal Interaction}
\label{subsec::NIF}

For any link $(i,j)$, we define the \emph{angular error} $\theta_{ij}^{\mathrm{err}} \in \left]- \frac{\pi}{L}, \frac{\pi}{L}\right]$ as the difference between $\theta_{ij}$ and the closest multiple of $2\pi/L$ (see Fig.~\ref{fig::vectors}), that is,
\begin{align}\label{eq::theta_ij^err}
    \theta_{ij}^{\mathrm{err}} &\coloneqq \theta_{ij} - 
    \frac{2 \pi}{L}
    \arg \min_{q \in \mathbb{Z}} \left\{ \left\vert \theta_{ij} - q\frac{2 \pi}{L} \right\vert \right\},
\end{align}

Then, the normal control input $\vec{u}_{\text{n},i}$ in \eqref{eq::controlLaw} is defined as 
\begin{align}\label{eq::normalInput}
    \vec{u}_{\text{n},i} = G_{\text{n},i}\sum_{j \in \mathcal{A}_i} f_{\text{n}}(\theta_{ij}^{\mathrm{err}}) \frac{\vec{r}_{ij}^\perp}{\norm{\vec{r}_{ij}}},
\end{align}
where $G_{\text{n,i}} \in \mathbb{R}_{\geq 0}$ is the normal control gain.
Each of these actions is applied in the direction of $\vec{r}_{ij}^\perp$, that is the vector normal to $\vec{r}_{ij}$, obtained by applying a $\pi/2$ counterclockwise rotation (see Fig.~\ref{fig::vectors}). The magnitude and sign of these forces are determined by the \emph{normal interaction function} $f_{\text{n}}: \,\left]-\frac{\pi}{L},\frac{\pi}{L}\right] \rightarrow \left[-1, 1\right[$, given by
\begin{align}\label{eq::LinearFn}
    f_{\text{n}}(\theta_{ij}^{\mathrm{err}}) & =  -\frac{L}{\pi} \,  \theta_{ij}^{\mathrm{err}}.
\end{align}
$f_\mathrm{n}$ is portrayed in Fig. 3b. 

We remark that by rotating the axis with respect to which angles $\theta_{ij}$ are measured,  our algorithm allows to achieve triangular or square lattices with different orientations.

\section{Numerical validation}
\label{sec::results}

In this section, we assess the performance and the robustness of our proposed control algorithm \eqref{eq::controlLaw} through an extensive simulation campaign. The experimental validation of the strategy is later reported in Sec. \ref{sec::robotarium}.
First in Sec. \ref{subsec::tuning}, using a numerical optimisation procedure, we tune the control gains $G_{\text{r},i}$ and $G_{\text{n},i}$ in \eqref{eq::radInput} and \eqref{eq::normalInput}, as the performance of the controlled swarm strongly depends on these values. 
Then in Sec. \ref{sec::robustness_analysis}, we assess the robustness of the control law with respect to (i) agents' failure and to (ii) noise, (iii) flexibility to pattern change, and (iv) scalability.
Finally in Sec. \ref{subsec::comparison}, we present a comparative analysis of our distributed control strategy and other approaches previously presented in the literature.
The simulations and experiments performed in this and the next Sections are summarised in Table \ref{tab:simulations}.

\begin{table}[t]
\begin{center}
    \caption{Simulations and experiments}
    \label{tab:simulations}
    \begin{tabular}{@{}lll@{}}
    \toprule
    Scenario & Section & Figure(s)  \\
    \midrule
    \emph{Control law \eqref{eq::controlLaw},\eqref{eq::radInput},\eqref{eq::normalInput}}\\
    \ \ Tuning & \ref{subsec::tuning} & \ref{fig::tuningSquaresTriangles}\\
    \ \ Validation & \ref{subsec::tuning} & \ref{fig::SimOverview}\\
    \ \ Robustness to faults & \ref{subsec::agents_removal} & \ref{fig::AgentsRemoval}\\
    \ \ Robustness to noise & \ref{subsec::noise} & \ref{fig::NoiseTest}\\
    \ \ Flexibility & \ref{subsec::DynLatt} & \ref{fig::dynlattices}\\
    \ \ Scalability & \ref{subsec::Scalability} & \ref{fig::Scalability}\\
    \ \ Comparison with \cite{Spears1999, Spears2004} & \ref{subsec::comparison} & \ref{fig::tuningSpears}, \ref{fig::scalabilitySpears}\\
    \hline
    \emph{Adaptive gain tuning} \\ \eqref{eq::controlLaw},\eqref{eq::radInput},\eqref{eq::normalInput},\eqref{eq::avgAngError} \\
    \ \ Validation & \ref{sec::adaptive} & \ref{fig::AdaptiveSquares}\\
    \ \ Robustness to faults & \ref{subsubsec::AdaptiveAgentsRemoval} & \ref{fig::AgentsRemovalAdaptive}\\
    \ \ Flexibility & \ref{subsubsec::AdaptiveFlexibility} & \ref{fig::AdaptiveDynLattices}\\
    \ \ Scalability & \ref{subsubsec::AdaptiveScalability} & \ref{fig::ScalabilityTestAdaptive}\\
    \hline
    Robotarium experiment & \ref{sec::robotarium} & \ref{fig::Robotarium}\\
    \bottomrule
\end{tabular}
\end{center}
\end{table}

\subsection{Simulation setup}
\label{sec:simulation_setup}
We consider a swarm consisting of $N = 100$ agents (unless specified differently).
To represent the fact that the agents are deployed from a unique source (as typically done in the literature \cite{Spears1999, Spears2004}), their initial positions are drawn randomly with uniform distribution from a disk of radius $r=2$ centred at the origin.
\footnote{That is, denoting with $U([a,b])$ the uniform distribution on the interval $[a,b]$, the initial position of each agent in polar coordinates $\vec{x}_i(0) \coloneqq (d_i,\phi_i)$ is obtained by independently sampling $\phi_i \sim U\left([0,2\pi[\right)$ and $d_i$ is chosen according to the  probability density function $p_\mathrm{l}(\xi): [0,r]\mapsto \mathbb{R}_{\geq 0}$ defined as $p_l(\xi) = 2 \xi/r^2$. } 

Initially, for the sake of simplicity and to avoid the possibility of some agents becoming disconnected from the group, we assume that $R_{\text{s}}$ in \eqref{eq:neighbourhood} is large enough so that
\begin{equation}\label{eq:assumption_sensing_radius}
    \forall i \in \mathcal{S}, \forall t \in \mathbb{R}_{\ge 0}, \quad \mathcal N_i(t) = \mathcal{S} \setminus i;
\end{equation}
i.e., any agent can sense the relative position of all others.
Later, in Sec. \ref{sec::robustness_analysis}, we will drop this assumption and show the validity of our control strategy also for smaller values of $R_{\text{s}}$.
All simulation trials are conducted in {\sc Matlab}\footnote{The code is available at \url{https://github.com/diBernardoGroup/SwarmSimPublic}.}, integrating the agents' dynamics using the forward Euler method with a fixed time step $\Delta t > 0$.
Moreover, the speed of the agents is limited to $V_{\mathrm{max}} > 0$.
The values of the parameters used in the simulations are reported in Tab.~\ref{tab:parameters}.
\begin{table}[t]
\begin{center}
    \caption{Simulation parameters}
    \label{tab:parameters}
    \begin{tabular}{@{}lll@{}}
    \toprule
    Parameter & Description & Value \\
    \midrule
    $R$ & Desired link length &  $1 \, \text{m}$ \\
    $R_{\min}$ & Minimum link length & $0.6 \, \text{m}$\\
    $R_{\max}$ & Maximum link length &  $1.1 \, \text{m}$ \\
    $V_{\max}$ & Maximum speed & $5 \, \text{m/s}$ \\
    $t_{\max}$ & Maximum simulation time & $200$ s\\
    $\Delta t$ & Integration step & $0.01 \, \text{s}$\\
    $T_{\text{w}}$ & Time window & $10 \, \text{s}$ \\
    $a$ & Radial interaction function $f_{\text{r}}(\cdot)$ & 0.15 \\
    $b$ & '' & 0.15\\
    $c$ & '' & 5\\    
    \bottomrule
\end{tabular}
\end{center}
\end{table}

\subsubsection*{Performance evaluation}

To assess the performance of the controlled swarm we exploit the metrics $e_{\theta}$ and $e_L$ given in Definitions \ref{def::regularity} and \ref{def::compactness}.
Namely, we select empirically the thresholds $e_{\theta}^* = 0.2$ and $e_L^* = 0.3$, which are associated to satisfactory compactness and regularity of the swarm.
Then, letting $T_{\text{w}}>0$ be the length of a time window, we say that $e_{\theta}$ is at \emph{steady-state} from time $t'=k \Delta t$ (for $k \in \mathbb{Z}$) if 
\begin{equation}
  \vert e_{\theta}(t') - e_{\theta}(t' - j\Delta t) \vert \leq 0.1 \, e_{\theta}^*, \quad 
  \forall j \in \left\{ 1, 2, \dots, \left\lfloor \frac{T_{\text{w}}}{\Delta t} \right\rfloor \right\}.
\end{equation}
We give an analogous definition for the steady state of $e_L$ (using $e_L^*$ rather than $e_\theta^*$).
Then, we say that in a trial the swarm \emph{achieved steady-state} at time $t_{\mathrm{ss}}$ if there exists a time instant such that both $e_{\theta}$ and $e_L$ are at steady state, and $t_{\mathrm{ss}}$ is the smallest of such time instants.
Moreover, we deem the trial \emph{successful} if $e_{\theta}(t_{\mathrm{ss}}) < e_{\theta}^*$ and $e_L(t_{\mathrm{ss}}) < e_L^*$.
If in a trial steady-state is not reached in the time interval $[0, t_{\max}]$, the trial is stopped (and deemed unsuccessful).
We define
\begin{equation}
    e_{\theta}^{\mathrm{ss}} \coloneqq
    \begin{dcases}
         e_{\theta}(t_{\mathrm{ss}}), & \text{if steady state was achieved}, \\
         e_{\theta}(t_{\mathrm{max}}), & \text{otherwise}. \\
    \end{dcases}
\end{equation}
\begin{equation}
    e_L^{\mathrm{ss}} \coloneqq
    \begin{dcases}
         e_L(t_{\mathrm{ss}}), & \text{if steady state was achieved}, \\
         e_L(t_{\mathrm{max}}), & \text{otherwise}. \\
    \end{dcases}
\end{equation}

Finally, to asses how quickly the pattern if formed, we define
\begin{align}
    T_{\theta} &\coloneqq\min \{t' \in \mathbb{R}_{\geq 0} :\ e_{\theta}(t') \leq e_{\theta}^*, \;\; \forall t \geq t'\}, \\
    T_L &\coloneqq\min \{t'' \in \mathbb{R}_{\geq 0} :\ e_L(t'') \leq e_L^*, \;\; \forall t \geq t'' \}.
\end{align} 

\subsection{Tuning of the control gains}
\label{subsec::tuning}
\begin{figure}[t!]
    \centering
    \includegraphics[width=1\columnwidth]{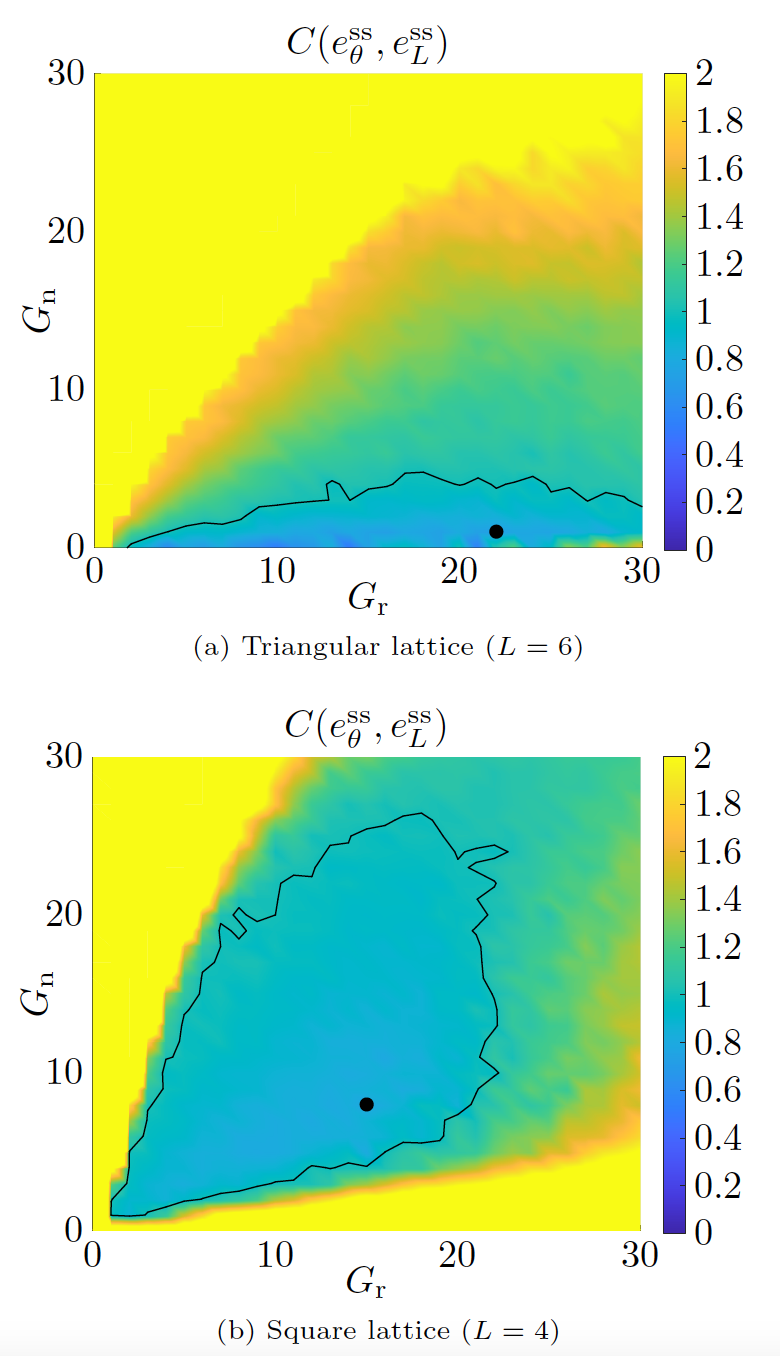}
    \caption{Tuning of the control gains $G_{\text{r}}$ and $G_{\text{n}}$ with (a) $L=6$ and (b) $L=4$ (§ \ref{subsec::tuning}).
    The black dots correspond to $(G_{\text{r}}^*,G_{\text{n}}^*)_{L=6}$ and $(G_{\text{r}}^*,G_{\text{n}}^*)_{L=4}$, minimising $C$. 
    The black curves delimit the regions where $C \leq 1$.
    }
    \label{fig::tuningSquaresTriangles}
\end{figure}

For the sake of simplicity, in this section we assume that 
$G_{\text{r},i} = G_{\text{r}}$ and $G_{\text{n},i} = G_{\text{n}}$, for all $i \in \mathcal{S}$; later, in Sec. \ref{sec::adaptive}, we will present an adaptive control strategy allowing each agent to independently vary online its own control gains.
To select the values of $G_{\text{r}}$ and $G_{\text{n}}$ giving the best performance in terms of regularity and compactness, we conducted an extensive simulation campaign and evaluated, for each pair $(G_{\text{r}}, G_{\text{n}}) \in \{0, 1, \dots ,30\} \times \{0, 1, \dots ,30\}$, the following cost function, averaged over $30$ trials, starting with random initial conditions:
\begin{equation}
    C(e_{\theta}^{\mathrm{ss}}, e_{L}^{\mathrm{ss}}) = 
    \left(\frac{e_{\theta}^{\mathrm{ss}}}{e_{\theta}^*}\right)^2 + \left(\frac{e_{L}^{\mathrm{ss}}}{e_{L}^*}\right)^2.
    \label{eq::costFunction}
\end{equation}
The results are reported in Fig.~\ref{fig::tuningSquaresTriangles} for the triangular ($L=6$) and the square ($L=4$) lattices; in the former case, the pair $(G_{\text{r}}^*, G_{\text{n}}^*)_{L=4}$ minimising $C$ is $(22, 1)$, whereas in the latter case it is $(G_{\text{r}}^*, G_{\text{n}}^*)_{L=6} = (15, 8)$.
Both pairs achieve $C\le 1$, implying $e_\theta^{\mathrm{ss}} \le e^*_\theta$ and $e_L^{\mathrm{ss}} \le e_L^*$.

\begin{figure*}[t!]
    \centering
    \includegraphics[width=2\columnwidth]{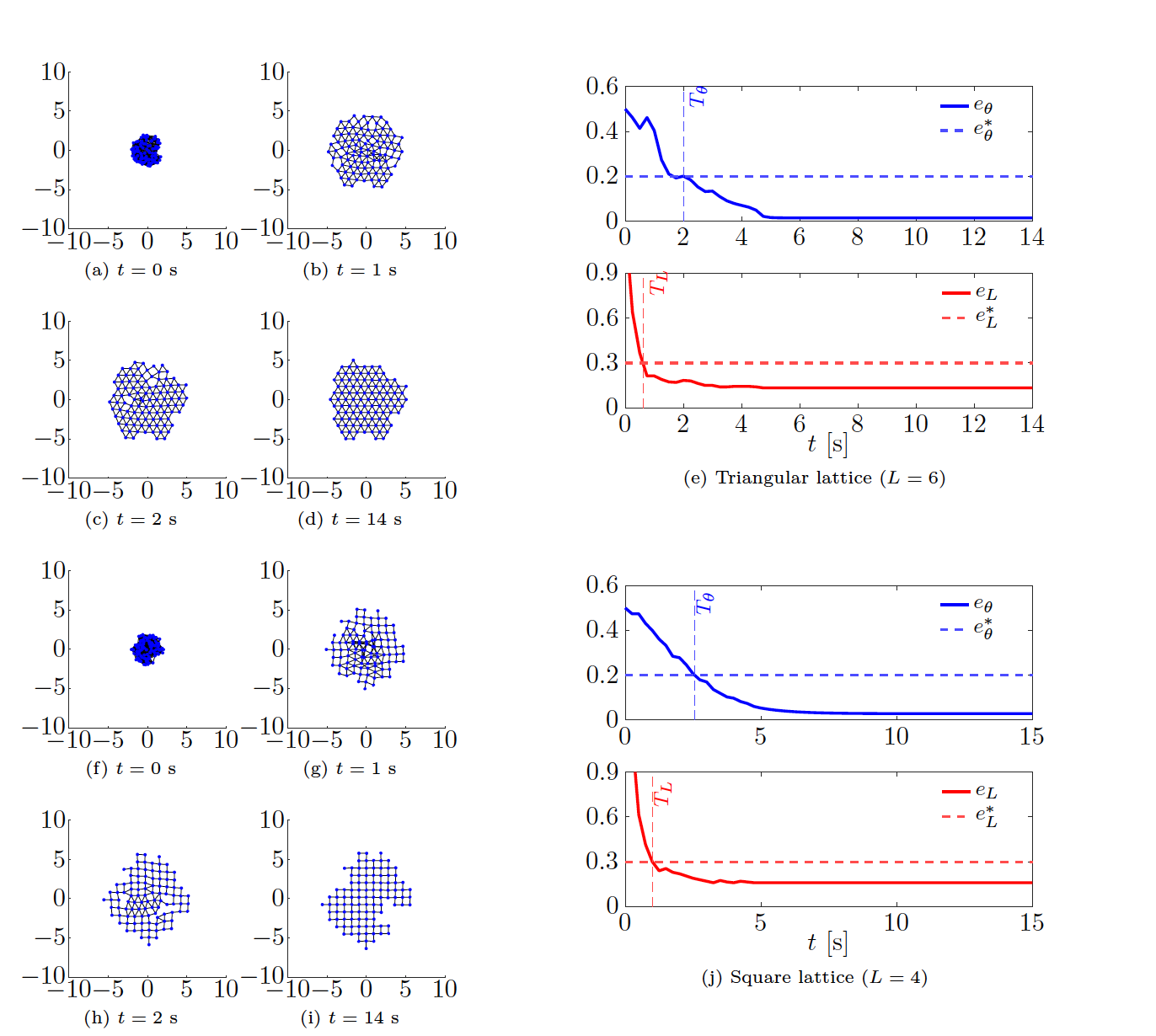}
    \caption{
    Snapshots at different time instants of a swarm forming (a)--(d) a triangular lattice (f)--(i) and a square lattice (§ \ref{subsec::tuning}). 
    Panels e and j show the time evolution of the metrics $e_{\theta}$ and $e_L$ for the cases that $L=6$ and $L=4$, respectively.
    When $L=6$, we set $(G_{\text{r}},G_{\text{n}}) = (G_{\text{r}}^*,G_{\text{n}}^*)_{L=6}$; when $L=4$, we set $(G_{\text{r}},G_{\text{n}}) = (G_{\text{r}}^*,G_{\text{n}}^*)_{L=4}$.
    }
    \label{fig::SimOverview}
\end{figure*}

In Fig.~\ref{fig::SimOverview}, we report three snapshots at different time instants of two representative simulations, together with the metrics $e_{\theta}(t)$ and $e_L(t)$, for the cases of a triangular and a square lattice, respectively.
The control gains were set to the optimal values $(G_{\text{r}}^*,G_{\text{n}}^*)_{L=6}$ and $(G_{\text{r}}^*,G_{\text{n}}^*)_{L=4}$.
In both cases, the metrics quickly converge below their prescribed thresholds, as $\max \{ T_\theta,$ $T_L\} < 2.75 \, \text{s}$.
Finally, note that $e_L(t)$ decreases faster than $e_{\theta}(t)$, meaning that the swarm tends to first reach the desired level of compactness and then agents' positions are rearranged to achieve the desired pattern.

\subsection{Robustness analysis}
\label{sec::robustness_analysis}

In this section, we investigate numerically the properties that we required in Sec.~\ref{sec::ProblemStatement}, that is robustness to faults and noise, flexibility, and scalability.

\subsubsection{Robustness to faults}
\label{subsec::agents_removal}
\begin{figure}[t!]
    \centering
    \includegraphics[width=1\columnwidth]{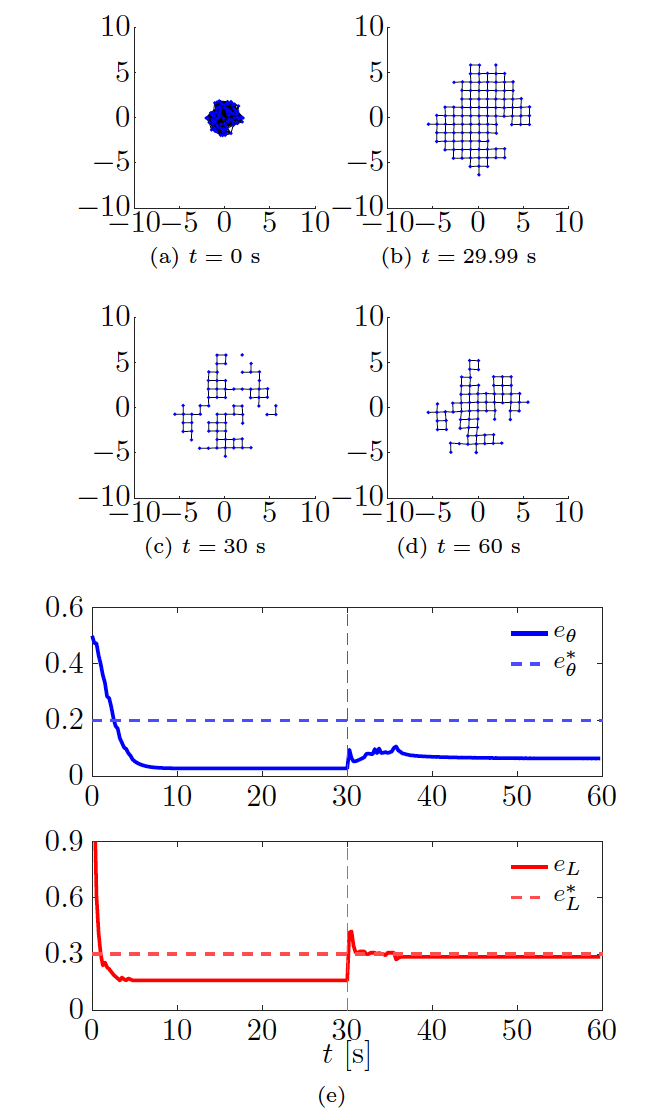}
    \caption{Robustness to removal of agents (§ \ref{subsec::agents_removal}).
    Panels a--d show snapshots at different time instants. Panel e shows the time evolution of the metrics; dashed vertical lines denote the time instant when agents are removed. 
    $L=4$, $(G_{\text{r}},G_{\text{n}}) = (G_{\text{r}}^*,G_{\text{n}}^*)_{L=4}$.
    }
    \label{fig::AgentsRemoval}
\end{figure}

We ran a series of simulations in which we removed a percentage of the agents at a certain time instant, and assessed the capability of the swarm to recover the desired pattern.
For the sake of brevity, we report one of them in Fig.~\ref{fig::AgentsRemoval}, where, with $L=4$, 30\% of the agents were removed at random at time $t = 30 \, \mathrm{s}$.
We notice that, as the agents are removed, $e_L(t)$ and $e_{\theta}(t)$ suddenly increase, but, after a short time, they converge again to values below the thresholds, recovering the desired pattern, despite the formation of small holes that increase $e_L^{\mathrm{ss}}$.

\subsubsection{Robustness to noise}
\label{subsec::noise}
\begin{figure}[t!]
    \centering
    \includegraphics[width=1\columnwidth]{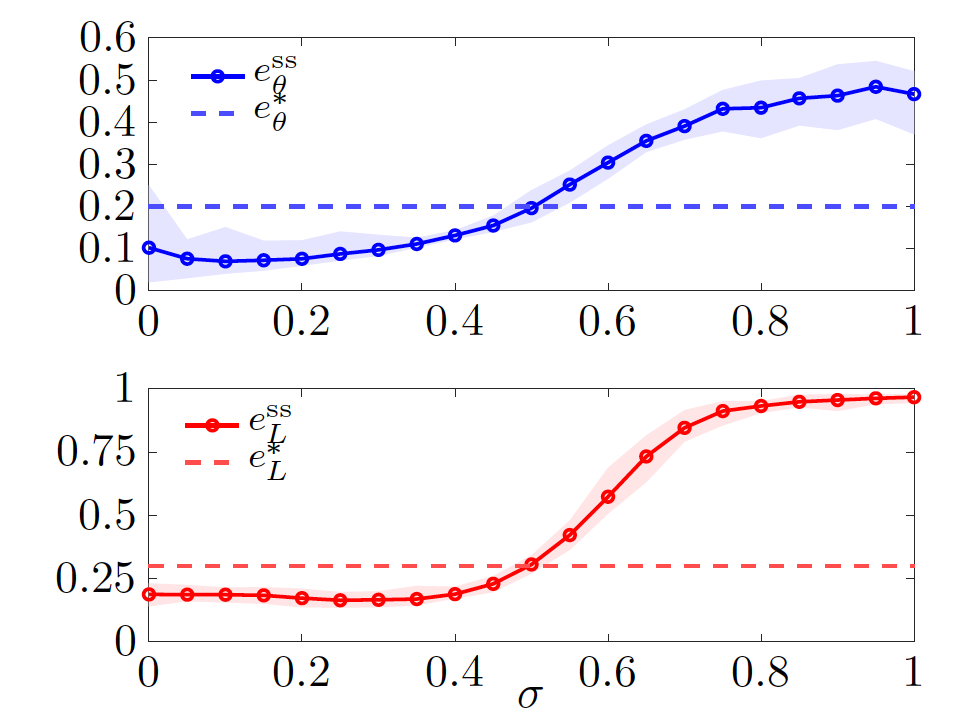}
    \caption{Robustness to noise (§ \ref{subsec::noise}).
    $e_L^\mathrm{ss}$ e $e_\theta^\mathrm{ss}$, averaged over $M = 30$ trials, varying the standard deviation of the Gaussian noise.
    The shaded areas represent the maximum and minimum values obtained over the $M$ trials.
    $L=4$, $(G_{\text{r}},G_{\text{n}}) = (G_{\text{r}}^*,G_{\text{n}}^*)_{L=4}$.}
    \label{fig::NoiseTest}
\end{figure}

%
We assumed that the dynamics \eqref{eq::firstOrdDynamics} of each agent is affected by additive white Gaussian noise with standard deviation $\sigma$.
Then, we set $L=4$ and varied $\sigma$ in the interval $[0, 1]$ with increments of $0.05$. 
For each value of $\sigma$, we ran $M = 30$ trials, starting from random initial conditions, and report the average values of $e_{\theta}^{\mathrm{ss}}$ and $e_{L}^{\mathrm{ss}}$ in Fig.~\ref{fig::NoiseTest}.
We observe that large intensities of noise ($\sigma \geq 0.4$) worsen performance, up to the point of making the trials unsuccessful and preventing the swarm from forming the desired lattice.
Interestingly, smaller noise ($0 < \sigma \leq 0.2$) actually improves performance.
This is because small random displacements can prevent the agents from getting stuck in undesired configurations, including those containing holes.

\subsubsection{Flexibility}
\label{subsec::DynLatt}
\begin{figure}[t!]
    \centering
    \includegraphics[width=0.9\columnwidth]{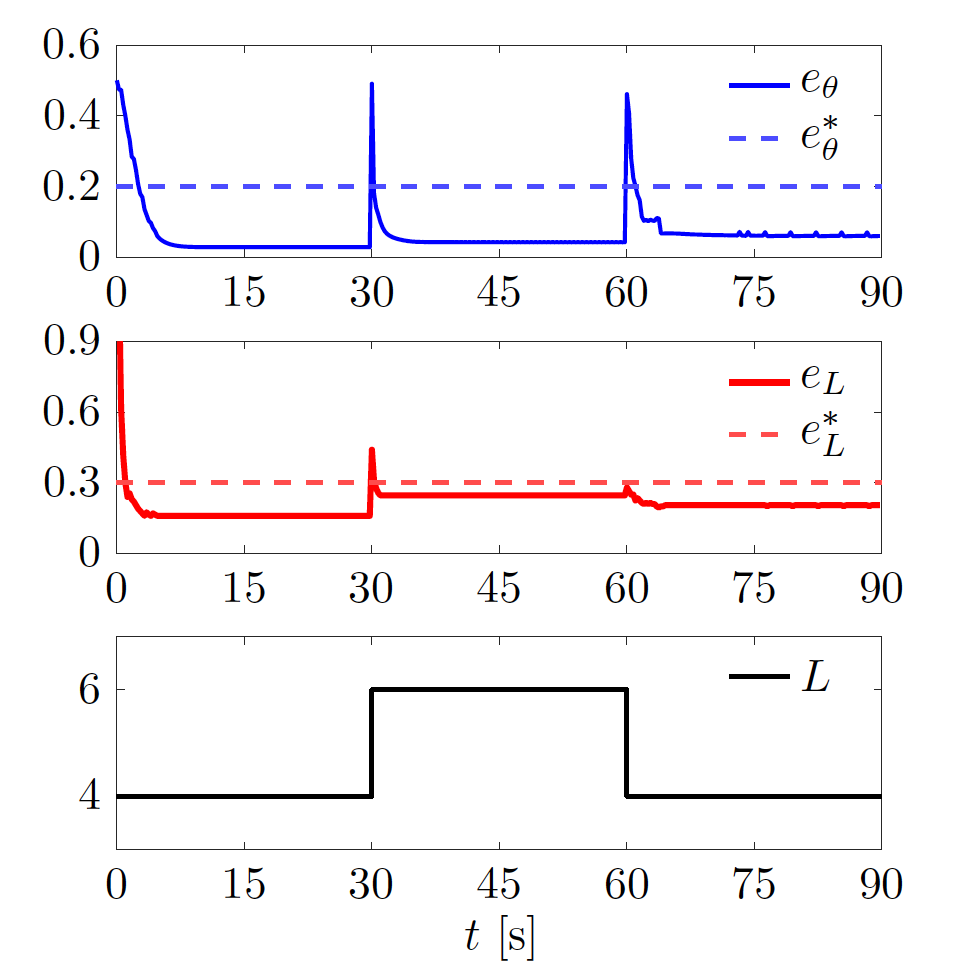}
    \caption{Flexibility to spatial reorganisation (§ \ref{subsec::DynLatt}).
    Time evolution of the metrics as $L$ changes.
    $(G_{\text{r}},G_{\text{n}}) = (G_{\text{r}}^*,G_{\text{n}}^*)_{L=4}$.
    }
    \label{fig::dynlattices}
\end{figure}

In Fig.~\ref{fig::dynlattices}, we report a simulation where $L$ was initially set equal to $4$ (square lattice), changed to $6$ (triangular lattice) at time $t=30  \ \mathrm{s}$, and finally changed back to $4$ at $t=60 \ \mathrm{s}$.
The control gains are set to $(G_{\text{r}}^*,G_{\text{n}}^*)_{L=4}$ and kept constant during the entire the simulation.
Clearly, as $L$ is changed, both $e_L$ and $e_{\theta}$ suddenly increase, but the swarm is quickly able to reorganise and reduce them below their prescribed thresholds in less than $5\, \mathrm{s}$, thus achieving the desired pattern.

\subsubsection{Scalability}
\label{subsec::Scalability}
Before properly testing for scalabiltiy, we dropped the assumption that \eqref{eq:assumption_sensing_radius} holds and characterised $e_L^{\mathrm{ss}}$ as a function of the sensing radius $R_{\text{s}}$.
The results are portrayed in Fig. 9a, showing that the performance starts deteriorating for approximately $R_{\text{s}} < 6 \, \text{m}$, until it becomes unacceptable for about $R_{\text{s}} < 1.1 \, \text{m}$.
Therefore, as a good trade-off between performance and feasibility, we set $R_{\text{s}} = 3 \, \text{m}$.
To test for scalability, we varied the number $N$ of agents (initially, $N = 100$), reporting the results in Fig. 9b.
We see that (i) the controlled swarm correctly achieves the desired pattern for at least four-fold changes in the size of the swarm, and (ii) compactness ($e_L^{\mathrm{ss}}$) improves as $N$ increases.
\begin{figure}[t!]
    \centering
    \includegraphics[width=1\columnwidth]{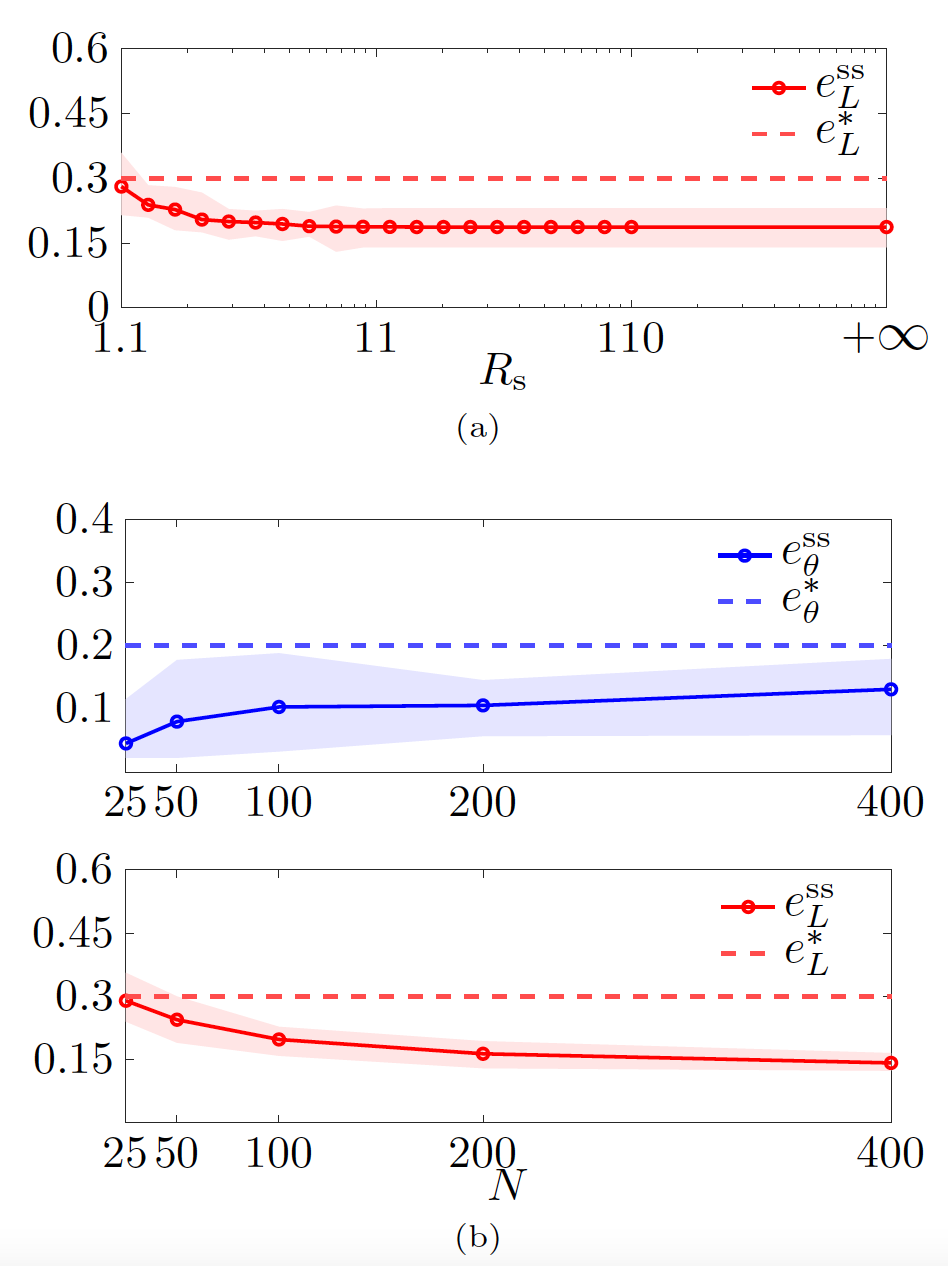}
    \caption{Scalability (§ \ref{subsec::Scalability}).
    (a) $e_L^\mathrm{ss}$ averaged over $M=30$ trials with varying $R_{\text{s}}$.
    (b) $e_\theta^\mathrm{ss}$ and $e_L^\mathrm{ss}$ averaged over the trials, with varying $N$;
    $R_{\text{s}} = 3\, \text{m}$; agents' initial positions are drawn with uniform distribution from a disk with radius $r=\sqrt{N/25}$.
    The shaded areas represent the maximum and minimum values over the $M$ trials.
    $L=4$, $(G_{\text{r}},G_{\text{n}}) = (G_{\text{r}}^*,G_{\text{n}}^*)_{L=4}$.
    }    
    \label{fig::Scalability}
\end{figure}

\subsection{Comparison with \cite{Spears1999, Spears2004}}
\label{subsec::comparison}
\begin{figure}[t!]
    \centering
    \includegraphics[width=1\columnwidth]{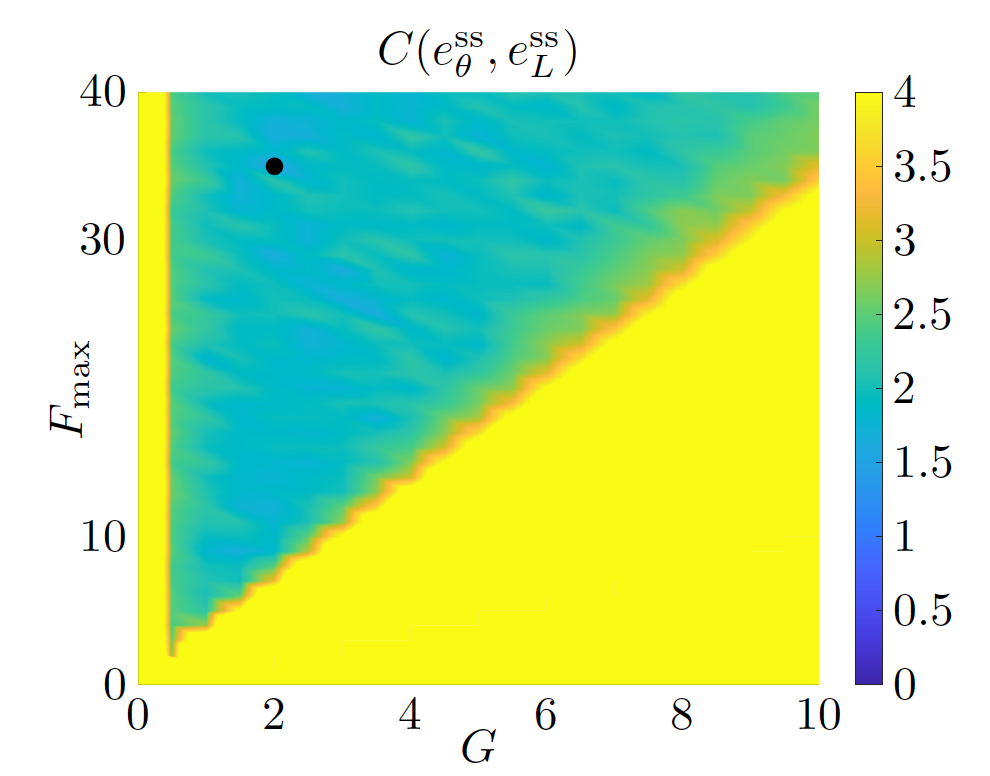}
    \caption{Tuning of parameters $G$ and $F_{\max}$ from \cite{Spears2004} (§ \ref{subsec::comparison}).
    The black dot denotes the optimal pair $(G^*,F_{\max}^*)$. $L=4$.
    }
    \label{fig::tuningSpears}
\end{figure}

We compared our control law \eqref{eq::controlLaw} to the so-called ``gravitational virtual forces strategy'' (see the Appendix) \cite{Spears1999,Spears2004}, that represent an established solution to geometric pattern formation problems. 
In \cite{Spears1999,Spears2004}, a second order damped dynamics is considered for the agents.
Hence, for the sake of comparison, we reduced that model to the first order model in \eqref{eq::firstOrdDynamics}, by assuming that the viscous friction force is significantly larger than the inertial one.

To select the gravitational gain $G$ and the saturation value $F_{\max}$ in the control law from \cite{Spears1999, Spears2004}, we applied the same tuning procedure described in Sec. \ref{subsec::tuning}.
In particular, we considered $(G, F_{\max}) \in \{0, 0.5, \dots, 10\} \times \{0, 1, \dots, 40\}$, and performed $30$ trials for each pair of parameters, obtaining as optimal pair for the square lattice $(G^*, F_{\max}^*) = (35, 2)$ (see Fig.~\ref{fig::tuningSpears}).
All other parameters where kept to the default values in Tab. \ref{tab:parameters}.

\begin{figure}[t!]
    \centering
    \includegraphics[width=1\columnwidth]{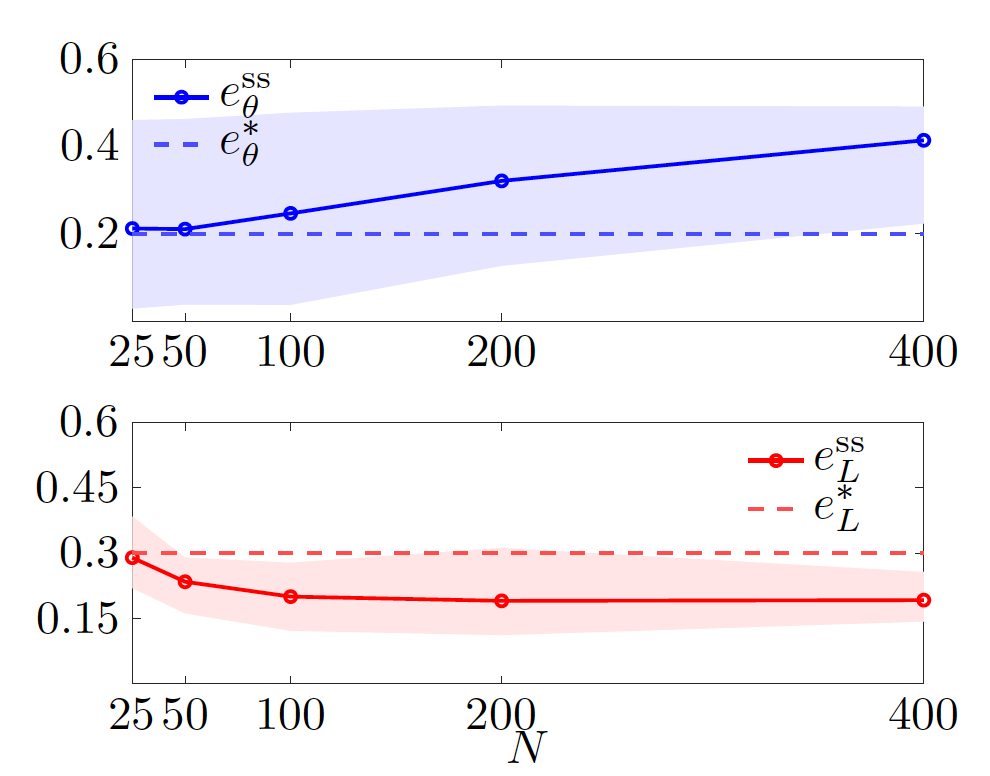}
    \caption{Scalability test for the algorithm from \cite{Spears2004} (§ \ref{subsec::comparison}).
    $e_L^\mathrm{ss}$ and  $e_\theta^\mathrm{ss}$ averaged over $M=30$ trials, as $N$ varies.
    Agents' initial positions are drawn with uniform distribution from a disk of radius $r=\sqrt{N/25}$.
    The shaded area represents the maximum and minimum values over the trials.
    $L=4$, $(G, F_{\max}) = (G^*, F_{\max}^*)$.
    }
    \label{fig::scalabilitySpears}
\end{figure}

Then, we performed the same scalability test in Sec. \ref{subsec::Scalability} and report the results in Fig. \ref{fig::scalabilitySpears}.
Remarkably, by comparing these results with those in the previous Fig. 9b, we see that our proposed control strategy performs better, obtaining much smaller values of $e_{\theta}^{\mathrm{ss}}$, regardless of the size $N$ of the swarm.
In particular, the control law from \cite{Spears1999, Spears2004} only rarely achieves $e_{\theta}^{\mathrm{ss}} \le e_{\theta}^*$, implying a low success rate. 

\section{Adaptive tuning of control gains}
\label{sec::adaptive}

Tuning the control gains (here $G_{\text{r}, i}$ and $G_{\text{n}, i}$) can in general be a tedious and time-consuming procedure.
Therefore, to avoid it, we propose the use of a simple adaptive control law, that might also improve the robustness and flexibility of the swarm.
Specifically, for the sake of simplicity, $G_{\text{r},i}$ is set to a constant value $G_{\text{r}}$ for all the swarm, while each agent computes its gain $G_{\text{n},i}$ independently, using only local information.
Letting $e_{\theta,i} \in [0,1]$ be the \emph{average angular error} for agent $i$, given by
\begin{equation}
    e_{\theta,i}\coloneqq\frac{L}{\pi} \frac{1}{\vert \mathcal{A}_i \vert} \sum_{j\in \mathcal{A}_i} \vert \theta_{ij}^{\mathrm{err}} \vert,
    \label{eq::avgAngError}    
\end{equation}
$G_{\text{n},i}$ is varied according to the law
\begin{subequations}
\begin{align}
    \frac{\mathrm{d}}{\mathrm{d}t}{G}_{n,i}(t)&=\begin{cases}
    \alpha \, (e_{\theta,i}(t)-e_{\theta}^*), &\mbox{if } e_{\theta,i}(t) > e_{\theta}^*, \\
    0, &\text{otherwise}.
    \end{cases} \label{eq::adaptationLaw:1}\\
    G_{\text{n},i}(0)&=0,
    \label{eq::adaptationLaw:2}
\end{align}
\label{eq::adaptationLaw}
\end{subequations}
\noindent where $\alpha > 0$ is an adaptation gain and $e_{\theta}^*$ (introduced in §\ref{sec:simulation_setup}) is used to determine the amplitude of the dead-zone. 
Here, we empirically choose $\alpha=3$. 
To evaluate the effect of the adaptation law, we also define the average normal gain of the swarm $\bar{G}_{\text{n}}(t) \coloneqq \frac{1}{N} \sum_{i=1}^N G_{\text{n},i}(t)$.

\begin{figure}[t!]
    \centering
    \includegraphics[width=1\columnwidth]{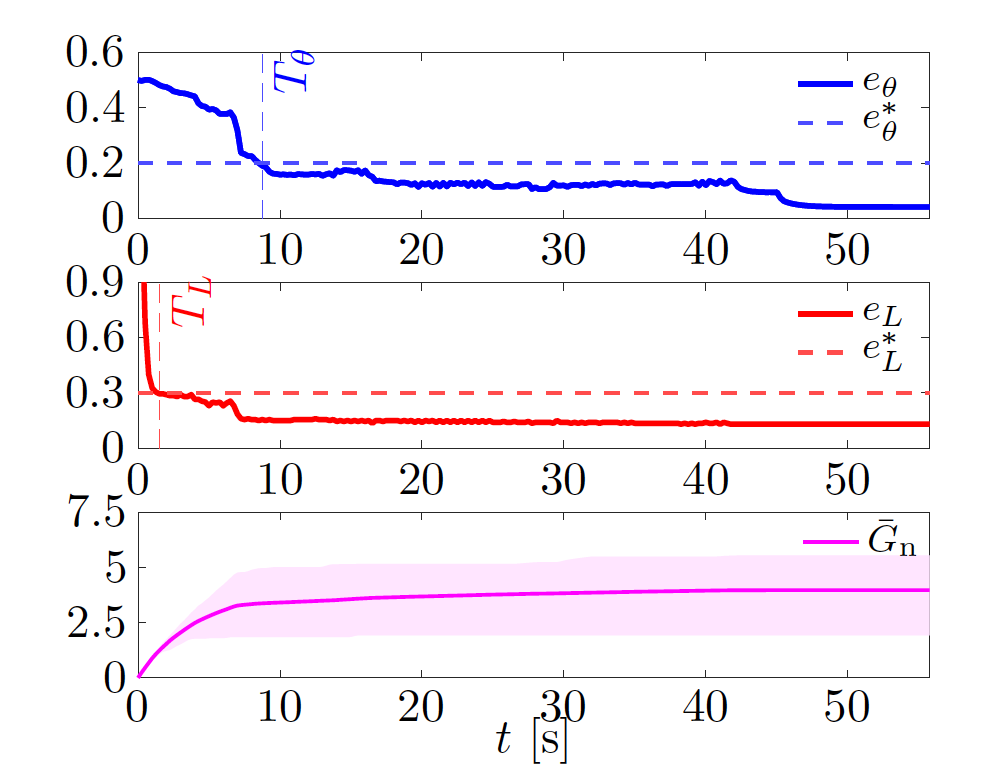}
    \caption{Pattern formation using the adaptive tuning law \eqref{eq::adaptationLaw} (§ \ref{sec::adaptive}).
    Initial conditions are the same as those of the simulation in Fig.~\ref{fig::SimOverview}.
    The shaded magenta area is delimited by $\max_{i \in \mathcal{S}} G_{\mathrm{n}, i}$ and $\min_{i \in \mathcal{S}} G_{\mathrm{n}, i}$. $L=4$, $G_{\text{r}}=15$.
    }
    \label{fig::AdaptiveSquares}
\end{figure}

In Fig. \ref{fig::AdaptiveSquares}, we report the time evolution of $e_L$, $e_{\theta}$, and of $\bar{G}_{\text{n}}$ for a representative simulation. 
First, we notice that the average normal gain $\bar{G}_{\text{n}}$ eventually settles to a constant value.
Moreover, comparing the results with the case in which the gains $G_{\mathrm{n},i}$ are not chosen adaptively (see Sec \ref{subsec::tuning} and Fig. 5j, here $T_\theta$, $T_L$ and $t_\mathrm{ss}$ are larger (meaning longer convergence time) but $e_\theta^\mathrm{ss}$ and $e_L^\mathrm{ss}$ are smaller (meaning better regularity and compactness performance).

\subsection{Robustness analysis}

Next, we test robustness to faults, flexibility, and scalability for the adaptive law \eqref{eq::adaptationLaw}, similarly to what we did in Sec. \ref{sec::robustness_analysis}.

\subsubsection{Robustness to faults}
\label{subsubsec::AdaptiveAgentsRemoval}

\begin{figure}[t!]
    \centering
    \includegraphics[width=1\columnwidth]{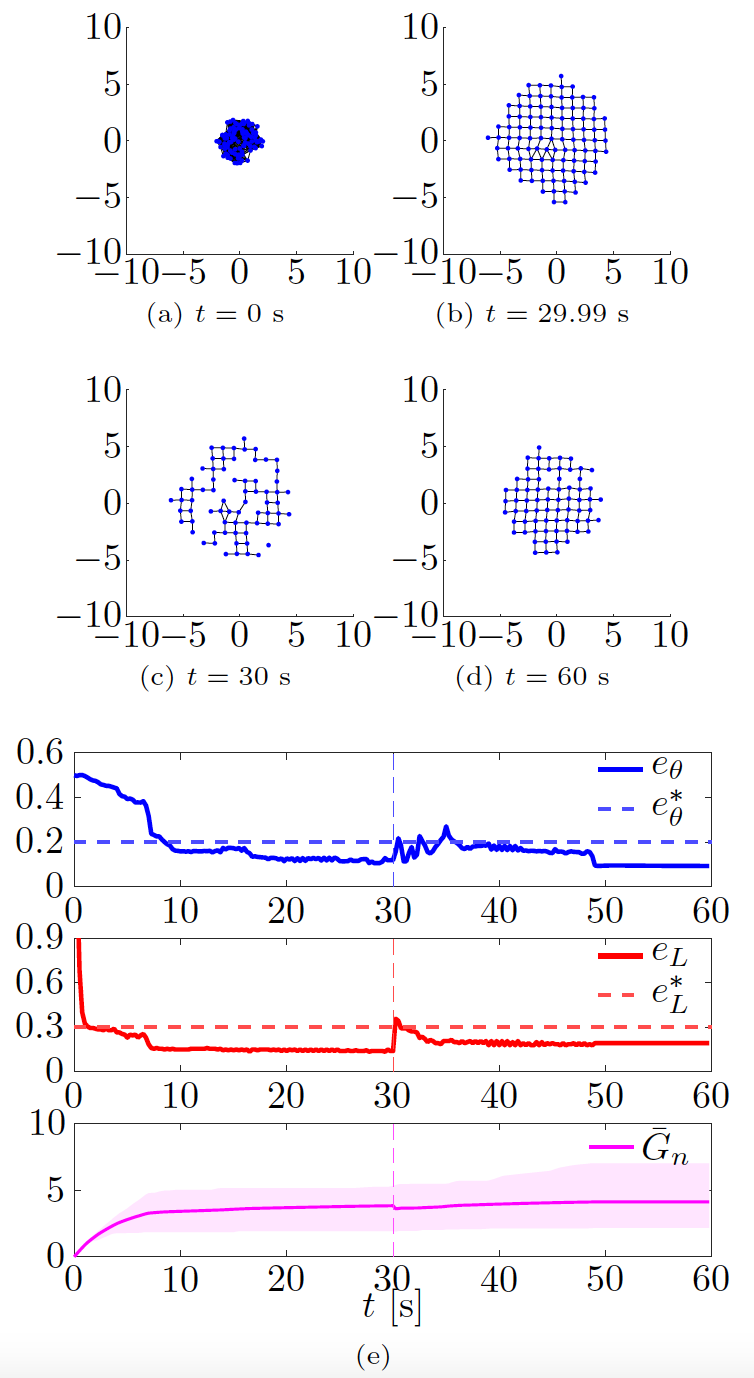}
    \caption{Robustness to agents removal using the adaptive tuning law \eqref{eq::adaptationLaw} (§ \ref{subsubsec::AdaptiveAgentsRemoval}). Initial conditions are the same as those of the simulation in Fig.~\ref{fig::AgentsRemoval}.
    Panels a--d show snapshots at different time instants. Panel e shows the time evolution of the metrics; dashed vertical lines denote the time instant when agents are removed. 
    The shaded magenta area is delimited by $\max_{i \in \mathcal{S}} G_{\mathrm{n}, i}$ and $\min_{i \in \mathcal{S}} G_{\mathrm{n}, i}$.
    $L=4$, $G_{\text{r}}=15$.
    }
    \label{fig::AgentsRemovalAdaptive}
\end{figure}

We ran a series of agent removal tests, as described in Sec. \ref{subsec::agents_removal}. 
For the sake of brevity, we report the results of one of such tests with $L=4$ in Fig. \ref{fig::AgentsRemovalAdaptive}.
At $t=30$ s, 30\%  of  the  agents are  removed; yet, after  a short time the swarm reaggregates to recover  the  desired  lattice. 

\subsubsection{Flexibility}
\label{subsubsec::AdaptiveFlexibility}

\begin{figure}[t!]
    \centering
    \includegraphics[width=0.9\columnwidth]{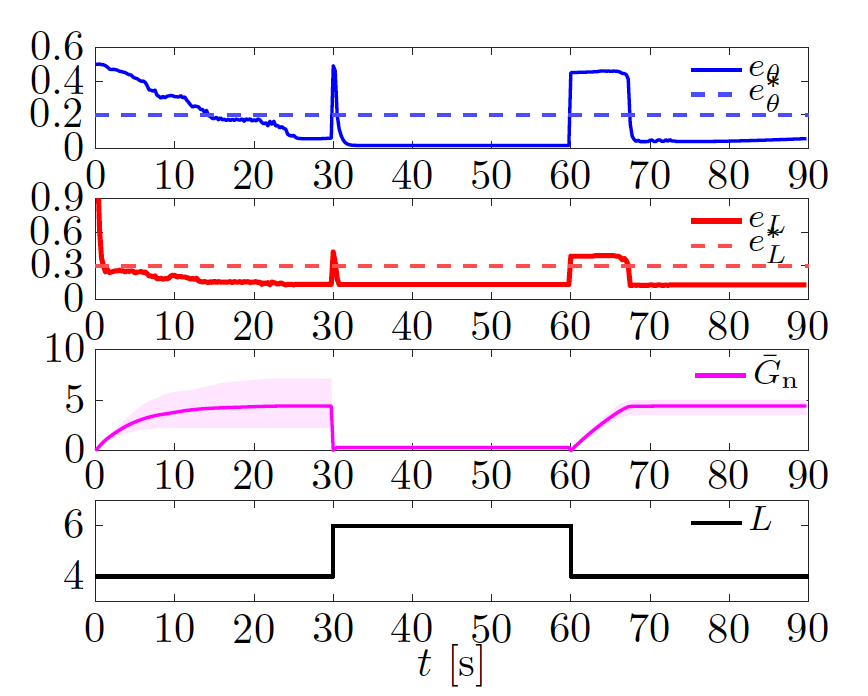}
    \caption{Flexibility using the adaptive tuning law \eqref{eq::adaptationLaw} (§ \ref{subsubsec::AdaptiveFlexibility}).
    Initial conditions are the same as those of the simulation in Fig.~\ref{fig::dynlattices}.
    The shaded magenta area is delimited by $\max_{i \in \mathcal{S}} G_{\mathrm{n}, i}$ and $\min_{i \in \mathcal{S}} G_{\mathrm{n}, i}$.
    }
    \label{fig::AdaptiveDynLattices}
\end{figure}

We repeated the test in Sec. \ref{subsec::DynLatt}, with the difference that this time we set $G_{\text{r}} = 18.5$ (that is the average between the optimal gain for square and triangular patterns), and set $G_{\text{n}, i}$  according to law \eqref{eq::adaptationLaw}, resetting all $G_{\mathrm{n}, i}$ to $0$ when $L$ is changed.
The results are shown in Fig.~\ref{fig::AdaptiveDynLattices}. 
When compared to the non-adaptive case (Fig. \ref{fig::dynlattices}), here $e_\theta^\mathrm{ss}$ and $e_L^\mathrm{ss}$ are smaller (better pattern formation), but $T_\theta$ and $T_L$ are larger (slower), especially when forming square patterns. 
Interestingly, when $L = 4$, $\bar{G}_{\mathrm{n}}$ settles to about 5, while when $L = 6$ it settles to about 0.3, a much smaller value.

\subsubsection{Scalability}
\label{subsubsec::AdaptiveScalability}

\begin{figure}[t!]
    \centering
    \includegraphics[width=0.9\columnwidth]{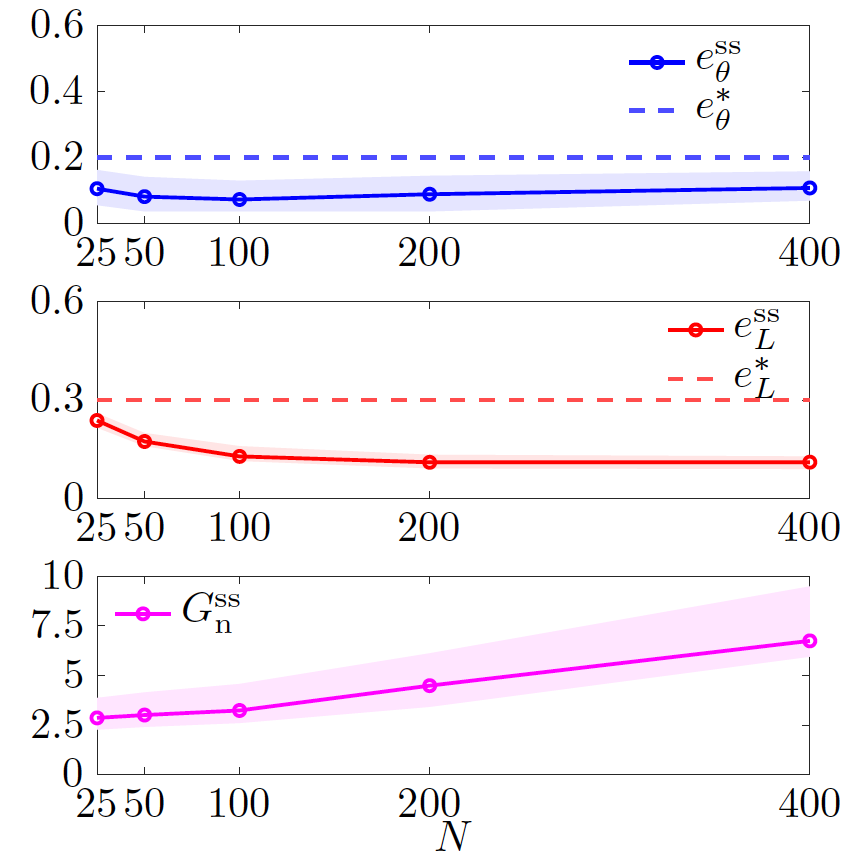}
    \caption{Scalability using the adaptive tuning law \eqref{eq::adaptationLaw} (§ \ref{subsubsec::AdaptiveScalability}). 
    $e_\theta^\mathrm{ss}$ and $e_L^\mathrm{ss}$ are averaged over $M=30$ trials with varying $N$.
    $R_{\text{s}} = 3\, \text{m}$; agents' initial positions are drawn with uniform distribution from a disk with radius $r=\sqrt{N/25}$.
    $G_{\text{n}}^{\text{ss}} \coloneqq \bar{G}_{\text{n}}(t_\mathrm{ss})$.
    The shaded areas represent the maximum and minimum values over the trials.
    $L=4$, $G_{\text{r}}=15$.
    }
    \label{fig::ScalabilityTestAdaptive}
\end{figure}

We repeated the test in Sec. \ref{subsec::Scalability}, setting again the sensing radius $R_{\text{s}}$ to 3 m and assessing performance while varying the size $N$ of the swarm; results are shown in Fig. \ref{fig::ScalabilityTestAdaptive}.
First, we notice that the larger the swarm is, the larger the steady state value of $\bar{G}_{\mathrm{n}}$ is.
Comparing the results with those of the static gains, in Fig. 9b, here we observe a slight improvement of performance, with a slightly smaller $e_\theta^{\mathrm{ss}}$. 

\section{Robotarium Experiments}
\label{sec::robotarium}

To further validate our control algorithm, we tested it in a real robotic scenario, using the open access \emph{Robotarium} platform \cite{Pickem2017, Mayya2020}.
The experimental setup features 20 differential drive robots (GRITSBot \cite{Pickem2015}), that can move in a 3.2 m $\times$ 2 m rectangular arena. The robots have a diameter of about 11 cm, a maximum (linear) speed of 20 cm/s, and a maximum rotational speed of about 3.6 rad/s.
To cope with the limited size of the arena, distances $\left\Vert \vec{r}_{ij} \right\Vert$ in \eqref{eq::Lennard-Jones} are doubled, while control inputs $\vec{u}_i$ are halved.
The Robotarium implementation includes a collision avoidance safety protocol and transforms the velocity inputs \eqref{eq::controlLaw} into appropriate acceleration control inputs for the robots.
Moreover, we run an initial routine to yield an initial condition in which the agents are aggregated as much as possible at the centre, similarly to what considered in Sec. \ref{sec::results}.

As a paradigmatic example, we performed a flexibility test (similarly to what done in Sec \ref{subsec::DynLatt} and reported in Fig. \ref{fig::dynlattices}).
During the first $33$ seconds, the agents reach an aggregated initial condition.
Then we set $L=4$ for $t \in [33, 165)$, $L=6$ for $t \in [165, 297)$, and $L=4$ for $t \in [297, 429]$, ending the simulation.
We used the static control law \eqref{eq::controlLaw}-\eqref{eq::radInput} and \eqref{eq::normalInput}, and to comply with the limited size of the arena, we scaled the control gains to the values $G_{\text{r}}=0.8$ and $G_{\text{n}}=0.4$, selected empirically.

\begin{figure}[h!]
    \centering
    \includegraphics[width=1\columnwidth]{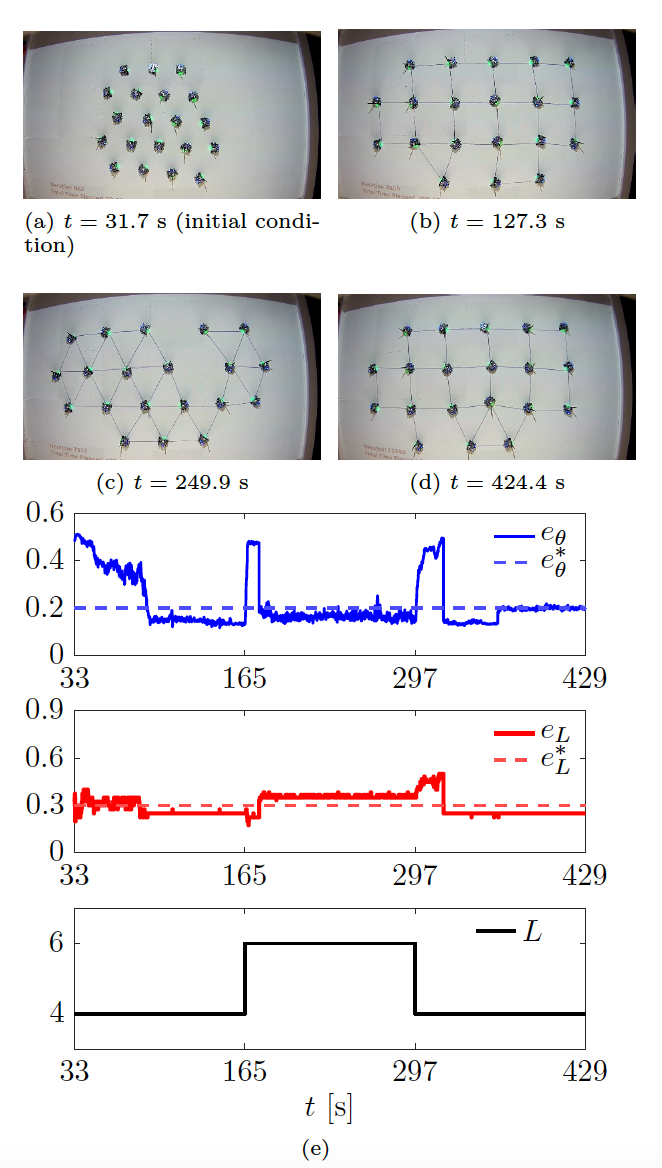}
    \caption{Flexibility test in Robotarium (§ \ref{sec::robotarium}). Panels a--d show the swarm at different time instants.
    Panel e shows the time evolution of the metrics and the parameter $L$.
    $(G_{\text{r}},G_{\text{n}}) = (G_{\text{r}}^*,G_{\text{n}}^*)_{L=4}$.
    }
    \label{fig::Robotarium}
\end{figure}

The resulting movie is available online (\url{https://github.com/diBernardoGroup/SwarmSimPublic}), while representative snapshots are reported in Fig. \ref{fig::Robotarium}, with the time evolution of the metrics.
The metrics qualitatively reproduce the behaviour obtained in simulation (see Fig. \ref{fig::dynlattices}). 
In particular, we obtain $e_{\theta}^\mathrm{ss} < e_{\theta}^*$, with both triangular and square patterns. 
On the other hand, we obtain $e_{L}^\mathrm{ss} < e_{L}^*$
when forming square patterns, but $e_{L}^\mathrm{ss} > e_{L}^*$ with triangular patterns; this is a consequence of the relatively small swarm size, and does not mean that the pattern is not achieved, as it can be seen in Fig. \ref{fig::Robotarium}.(c) showing the pattern is successfully achieved.

\section{Conclusions}
\label{sec::conclusions}

We presented a distributed control law to solve pattern formation for the case of square and triangular lattices, based on the use of virtual forces.
Our control strategy is distributed, only requires distance sensors and a compass, and does not need communication between the agents.
We showed via exhaustive simulations and experiments that the strategy is effective in achieving triangular and square lattice; we also compared it the distance-based strategy in \cite{Spears2004}, observing better performance particularly when the goal is that of achieving square lattices.
Additionally, we showed that the control law is robust to failures of the agents, to noise, is flexible to changes in the lattice and scalable with respect to the number of agents.
We also presented a simple yet effective adaptive law to automatically tune the gains so as to be able to switch the goal pattern in real-time.

In the future, we plan to study analytically the stability and convergence of the control law.
Additionally, we will investigate the extension to other patterns (e.g. hexagonal ones) and a more sophisticated adaptive law able to tune all the control gains at the same time.

\begin{appendix}

\section{The algorithm from \cite{Spears1999,Spears2004}}
\label{secA1}

Let us first introduce some useful notation. 
Given a real-valued function $x(t)$ and $a,b \in \mathbb{R}$ with $a<b$, we introduce the \emph{saturation} of $x(t)$ between $a$ and $b$, given by
\begin{equation*}
    \left[x(t) \right]_a^b \coloneqq \begin{cases}
    a, &\mbox{if} \;\; x(t)<a, \\
    x(t), &\mbox{if} \;\; a\leq x(t) \leq b,\\
    b, &\mbox{if} \;\; x(t)>b.
    \end{cases}
\end{equation*}

In \cite{Spears1999, Spears2004, Sailesh2014}, the agent dynamics is described by 
\begin{align}
\begin{cases}
\dot{\vec{x}}_i =\vec{v}_i, \\
\dot{\vec{v}}_i =\frac{1}{m}(\vec{u}_i-\mu \vec{v}_i),
\end{cases}
\quad \forall i \in \mathcal{S},
\label{eq::SecondOrderIntegrDamped}
\end{align}
where $\vec{u}_i\in \mathbb{R}^2$ is the control input, $m\in \mathbb{R}_{>0}$ is the mass of the agent and $\mu\in \mathbb{R}_{>0}$ is the friction damping factor. 
Recall that, as described in Sec. \ref{sec::ProblemStatement}, under a few assumptions, \eqref{eq::SecondOrderIntegrDamped} can be recast as \eqref{eq::firstOrdDynamics}.
The control input $\vec{u}_i$ is given by
\begin{equation}
    \vec{u}_i = \sum_{j=1}^N f(\Vert \vec{r}_{ij}\Vert) \frac{\vec{r}_{ij}}{\Vert\vec{r}_{ij}  \Vert}
    \label{eq::spearsContrInp},
\end{equation}
where $f$ is a gravitational-like virtual force, given by
\begin{equation}
    f(\Vert \vec{r}_{ij}\Vert )= \begin{cases}
   \left[ G \frac{m^2}{\Vert \vec{r}_{ij}\Vert ^2} \right]_0^{F_{\max}}, 
   &\mbox{if } \ 0\leq \Vert \vec{r}_{ij}\Vert \leq R,\\
    -\left[G \frac{m^2}{\Vert \vec{r}_{ij}\Vert ^2}\right]_0^{F_{\max}},
    &\mbox{if } \ R<\Vert \vec{r}_{ij}\Vert \leq1.5 R,\\
    0,
    &\mbox{otherwise}.
    \end{cases}
    \label{eq::spearsForce}
\end{equation}
In \eqref{eq::spearsForce}, $G, F_{\max} \in \mathbb{R}_{\geq0}$ are tunable control gains, and $R \in \mathbb{R}_{>0}$. 

The control law given by \eqref{eq::spearsContrInp} and  \eqref{eq::spearsForce} was showed to work for triangular lattices. 
To make it suitable for square patterns, a binary variable called \emph{spin} is introduced for each agent, and the swarm is divided in two subsets, depending on the value of their spin.
Then, agents with different spin aggregate at a distance of  $R$, while agents with the same spin do so at a distance of $\sqrt{2}R$. 
The extension to the case of hexagonal lattice is discussed in \cite{Sailesh2014} and requires communication among the agents.

\end{appendix}

\bibliographystyle{unsrt}
\bibliography{main}

\end{document}